\documentclass[12pt]{article}
\usepackage{amsmath, amssymb, amsfonts}
\usepackage[
  paperwidth=8.5in,
  paperheight=11in,
]{geometry}
\usepackage{hyperref}
\usepackage{booktabs}
\usepackage{float}
\usepackage{comment}
\usepackage[authoryear,round]{natbib}
\usepackage{xurl}
\usepackage{url}
\hypersetup{hidelinks}

\begin{document}

\title{Bayesian Inference for Incomplete $2 \times 2$ Diagnostic Tables}
\author{
Sara Antonijevic, Danielle Sitalo, and Brani Vidakovic}

\date{\small Department of Statistics, Texas A\&M University, College Station, TX 77843, USA}
\maketitle

\begin{abstract}
Incomplete reporting of diagnostic accuracy data remains a persistent problem in medical research. In many studies, only part of the $2 \times 2$ diagnostic table is reported, leaving denominators for diseased and non-diseased groups unknown and preventing direct calculation of sensitivity, specificity, predictive values, and related operating characteristics. To address this limitation, we develop hierarchical Bayesian models for reconstructing incomplete $2 \times 2$ diagnostic tables from such partial information. Two motivating scenarios are considered: one in which only a single test-outcome row is observed, and another in which true positives, false positives, and the total sample size are reported but the remaining cells are missing. The proposed models are illustrated on a benchmark breast MRI study with complete counts, treated as partially observed in order to assess reconstruction performance under controlled missingness. The framework yields posterior inference for the missing cell counts and associated diagnostic measures, together with uncertainty quantification in weakly identified settings.
\end{abstract}

\section*{Introduction}
\label{sec:intro}
Incomplete or partially reported diagnostic accuracy data remain a common obstacle to interpretation, reproducibility, and evidence synthesis. In many applied studies, authors report only fragments of the $2 \times 2$ diagnostic table, for example a single cell count, one observed row, or a pair of summary measures, without providing the full denominators for diseased and non-diseased groups. As a result, readers may be unable to reconstruct the full table, verify internal consistency, or derive clinically relevant quantities such as predictive values, false discovery rates, or expected error counts \citep{FDA2007,CochraneDTA2010}.

This problem is not merely clerical. Complete cross-tabulation is central to transparent reporting of diagnostic accuracy studies, and the STARD initiative was developed precisely to improve such reporting. The STARD 2015 revision explicitly recommends reporting both the cross-tabulation of index test results against the reference standard and a participant flow diagram showing how the study denominators were obtained \citep{Bossuyt2015,Cohen2016,EquatorSTARD}. When these elements are omitted, clinically important operating characteristics may no longer be recoverable from the published record.

Empirical assessments suggest that such omissions remain common. Earlier audits showed that incomplete reporting often prevented reconstruction of the full $2 \times 2$ table \citep{Smidt2005,Wilczynski2008}, and more recent work indicates that adherence to STARD recommendations remains uneven even in contemporary medical imaging diagnostic accuracy studies \citep{White2025}. For AI-centered diagnostic accuracy studies, the recent STARD-AI extension further underscores the need for clear reporting of dataset construction, model evaluation, and clinical applicability \citep{STARD_AI2025}. The practical consequence is that studies may be difficult to check, compare, or incorporate into evidence syntheses, even when the underlying clinical question is important.

A related methodological literature addresses situations in which the reference standard is applied only to a subset of participants, creating verification problems and the potential for work-up bias \citep{deGroot2011,BuzoianuKadane2008,UmemnekuChikere2019}. That literature is important for the present paper because one of our motivating examples arises from precisely such a setting. At the same time, incomplete reporting can also occur in studies where the reference standard is not obviously missing for design reasons, but the published article still omits sufficient cell counts to prevent recovery of the full diagnostic table. Our focus is on this reporting and reconstruction problem.

In practical terms, if only sensitivity and specificity are reported, a full $2 \times 2$ table cannot usually be reconstructed unless additional information is available, such as the total sample size, the number of diseased subjects, or an externally justified prevalence estimate. Because many evidence-synthesis frameworks require study-level $2 \times 2$ tables, missing denominators directly limit downstream meta-analytic use \citep{CochraneDTA2010}.

This paper develops statistical models for reconstructing incomplete $2 \times 2$ diagnostic tables from partially reported data. Our motivation comes from real examples in the diagnostic accuracy literature where key cells are unobserved in the published report. We consider two representative scenarios. In the first, based on \citet{Svirsky2002}, only the test-positive subgroup is reported in detail, leaving both cells of the test-negative row unobserved. This case is naturally connected to the literature on partial verification because the reference standard was applied only to a selected subgroup. In the second, based on \citet{Wismueller2020ICH}, counts for true positives and false positives are reported and the total sample size is known, but the negative-class counts are omitted. This second setting is better viewed as a constrained incomplete-table problem, since the known total sample size links the missing denominators.

Our aim is not to replace the broader verification-bias literature, nor to claim that incomplete diagnostic tables can always be uniquely recovered from sparse summaries alone. Rather, we develop Bayesian reconstruction strategies for settings in which deterministic recovery is impossible, but principled posterior inference on the missing denominators and derived operating characteristics remains feasible under clearly stated modeling assumptions.

The remainder of the paper is organized as follows. We first review notation for diagnostic $2 \times 2$ tables and summarize the binomial-$n$ problem that underlies our reconstruction strategy. We then present the two motivating incomplete-table scenarios, develop the corresponding models, and illustrate their performance on a benchmark example with complete counts treated as partially observed.

\section*{Review and Notation for $2 \times 2$ Diagnostic Accuracy Tables and the Binomial $n$ Problem}

A $2 \times 2$ diagnostic table, also called a confusion table, is the standard framework for summarizing the performance of an index test against a reference standard. It records the counts of true positives (TP), false positives (FP), false negatives (FN), and true negatives (TN), and serves as the basis for the usual measures of diagnostic performance.
\begin{table}[h!]
\centering
\begin{tabular}{lccr}
\hline
 & Disease Present & Disease Absent & Total \\
\hline
Test Positive & TP & FP & $n_+ = \mathrm{TP}+\mathrm{FP}$ \\
Test Negative & FN & TN & $n_- = \mathrm{FN}+\mathrm{TN}$ \\
\hline
Total & $n_1 = \mathrm{TP}+\mathrm{FN}$ & $n_2 = \mathrm{FP}+\mathrm{TN}$ & $N$ \\
\hline
\end{tabular}
\caption{Standard $2 \times 2$ diagnostic table showing true positives (TP), false positives (FP), false negatives (FN), and true negatives (TN), with row totals $n_+$, $n_-$, column totals $n_1$, $n_2$, and overall sample size $N$.}
\label{tab:diagnostic}
\end{table}

From these counts one obtains the familiar diagnostic accuracy measures:
\begin{alignat}{4}
\mathrm{Se} &= \frac{\mathrm{TP}}{\mathrm{TP}+\mathrm{FN}}, &\qquad
\mathrm{Sp} &= \frac{\mathrm{TN}}{\mathrm{FP}+\mathrm{TN}}, \\[6pt]
\mathrm{PPV} &= \frac{\mathrm{TP}}{\mathrm{TP}+\mathrm{FP}}, &\qquad
\mathrm{NPV} &= \frac{\mathrm{TN}}{\mathrm{FN}+\mathrm{TN}}, \\[6pt]
& & \mathrm{Accuracy} &= \frac{\mathrm{TP}+\mathrm{TN}}{N},
\end{alignat}
\citep{langlotz2003fundamental,Eusebi2013,Vidakovic2017}.

We write
$n_1 = \mathrm{TP}+\mathrm{FN}$
for the total number of diseased individuals,
$n_2 = \mathrm{FP}+\mathrm{TN}$
for the total number of non-diseased individuals, and
$N = n_1+n_2$
for the total sample size. When all four cell counts are available, all standard operating characteristics can be computed directly. When only partial information is reported, for example TP and FP without the corresponding denominators, the full table cannot be reconstructed from the published data alone. In that setting, sensitivity, specificity, and NPV are not identified without additional information or modeling assumptions.

\subsection*{Bayesian Approaches to the Binomial $n$ Problem}

A central ingredient in our reconstruction problem is the classical binomial-$n$ problem: infer the number of trials $n$ in a Bin$(n,p)$ model when $n$ is unknown and $p$ may also be unknown. This problem is well known to be difficult, especially when only a single observation, or very limited data, are available. Classical estimators based on moments or maximum likelihood can be unstable, particularly when the sample variance is close to the sample mean, and the difficulty becomes more pronounced when $p$ is small or $n$ is large \citep{Haldane1945,OlkinPetkau,DasGupta2005}.

Bayesian methods provide a natural way to regularize this problem by introducing prior information on both $n$ and $p$. Early work considered simple prior structures such as
$n \sim \mathrm{Uniform}(1,N)$
and
$p \sim \mathrm{Beta}(\alpha,\beta)$,
leading to posterior inference for $n$ through either posterior modes or posterior means under the chosen loss function \citep{DraperGuttman,CarrollLombard}. Rubin's empirical Bayes treatment of the problem was especially influential in motivating later hierarchical formulations \citep{Rubin1978}.

Subsequent developments introduced more flexible prior models. For example, \citet{Raftery1988} considered a predictive framework with
$n \sim \mathrm{Poisson}(\mu)$
and
$p \sim \mathrm{Uniform}(0,1)$.
Other extensions include beta priors for $p$ together with truncated Poisson priors for $n$ \citep{Bayoud2011}, as well as continuous gamma approximations for $n$ that simplify computation \citep{GunelChilko}. These approaches reduce the instability of purely classical procedures and allow substantive prior information about plausible ranges of $n$ and $p$ to enter the analysis.

A practically attractive alternative is the empirical Bayes or integrated-likelihood approach, which estimates $n$ jointly with beta hyperparameters for the prior on $p$ by maximizing the beta-binomial likelihood \citep{CarrollLombard,DasGupta2005}. Such methods are often computationally convenient and can perform well in small-sample settings where direct likelihood-based inference is erratic.

Because our incomplete-table problem involves missing denominators rather than merely missing cell probabilities, the binomial-$n$ literature provides a natural modeling foundation. In particular, the unknown stratum totals $n_1$ and $n_2$ can be viewed as latent trial counts that must be inferred from partially observed binomial information under suitable prior structure.

A recent and comprehensive review of estimation in the binomial-$n$ problem, including classical, Bayesian, empirical Bayes, and computational aspects, is given by \citet{GeorgievaVidakovic2025}.

\section*{Incomplete Diagnostic Tables}

Published diagnostic studies do not always report the full set of cell counts needed to reconstruct the $2 \times 2$ table. When only a subset of the cells is available, quantities such as sensitivity, specificity, negative predictive value, and overall accuracy may no longer be identified from the published data alone. In some cases, positive predictive value can still be computed from the reported true positives and false positives, but the absence of information on false negatives and true negatives prevents recovery of the complete diagnostic table.

The two examples below illustrate that incomplete $2 \times 2$ tables may arise in more than one way. In the first case, the missingness is linked to study design: only test-positive subjects underwent verification with the reference standard, so the test-negative row is unobserved. This setting is naturally connected to the literature on partial verification and work-up bias \citep{deGroot2011,BuzoianuKadane2008,VerificationBias,Kohn2022,UmemnekuChikere2019}. In the second case, the total sample size is known and the published report provides counts for true positives and false positives, but the negative-class counts are omitted. This is better viewed as an incomplete reporting problem with a known total, which leads to a different reconstruction strategy. Together, these two cases motivate the models developed in the next section.

\subsection*{Case 1: Partial Verification}

\citet{Svirsky2002} compared computer-assisted oral brush biopsy results with follow-up scalpel biopsy and histology in order to estimate the positive predictive value of an abnormal brush-biopsy result. Among 243 patients with abnormal brush-biopsy findings who then underwent scalpel biopsy, 93 were confirmed as dysplasia or carcinoma by histology and 150 were histology negative. Thus, PPV can be calculated directly within the test-positive subgroup as $93/243 \approx 0.38$.

The difficulty is that only patients with abnormal brush-biopsy results underwent the reference standard. Patients with normal brush-biopsy results were not verified by histology, so the entire test-negative row is unobserved. As a consequence, the numbers of false negatives and true negatives are unknown, and the full $2 \times 2$ table cannot be reconstructed from the published report. Sensitivity, specificity, negative predictive value, and overall accuracy therefore remain unidentified. In this sense, Case 1 is not merely an incomplete-table problem. It is also a partial-verification design, because the reference standard was applied only to a selected subgroup.

\begin{table}[h!]
\centering
\begin{tabular}{p{4.5cm}ccr}
\hline
 & Histology Positive & Histology Negative & Total \\
\hline
Brush biopsy abnormal (test positive) & 93 (TP) & 150 (FP) & 243 \\
Brush biopsy normal (test negative)   & ? (FN) & ? (TN) & ? \\
\hline
Total & $n_1 = ?$ & $n_2 = ?$ & $N = ?$ \\
\hline
\end{tabular}
\caption{Available and missing information from \citet{Svirsky2002}. Only patients with abnormal brush-biopsy results underwent scalpel biopsy and histology, so the test-negative row and all column and overall totals are unobserved.}
\label{tab:svirsky}
\end{table}

Although the system has at times been described in later discussions as AI-assisted, the technology in \citet{Svirsky2002} is more accurately viewed as an early rule-based computer-assisted diagnostic tool rather than artificial intelligence in the contemporary sense.

\subsection*{Case 2: Incomplete Reporting with Known $N$}

\citet{Wismueller2020ICH} evaluated an AI-based system for detecting intracranial hemorrhage on emergent head CT scans. The paper reports that 105 of 122 AI-positive cases were true positives, so PPV can be calculated as $105/122 \approx 0.86$. The study also reports the total number of scans, namely $N = 620$.

However, the total number of actual hemorrhage cases, $n_1 = \mathrm{TP} + \mathrm{FN}$, is not reported, and neither are the counts of false negatives and true negatives. Consequently, sensitivity and specificity cannot be computed directly, and the complete $2 \times 2$ table cannot be reconstructed from the published data alone.
\begin{table}[h!]
\centering
\begin{tabular}{lccr}
\hline
 & ICH Present & ICH Absent & Total \\
\hline
AI positive (flagged as ICH) & 105 (TP) & 17 (FP) & 122 \\
AI negative (not flagged)    & ? (FN) & ? (TN) & ? \\
\hline
Total & $n_1 = ?$ & $n_2 = ?$ & $N = 620$ \\
\hline
\end{tabular}
\caption{Partially observed $2 \times 2$ table implied by \citet{Wismueller2020ICH}. The publication reports the AI-positive counts and the total sample size $N=620$, but the negative-class counts and diseased/non-diseased totals remain unknown.}
\label{tab:wismueller}
\end{table}

Case 2 differs from Case 1 in an important way. Here the main obstacle is not selective verification of the reference standard, but incomplete reporting despite a known total sample size. Because $N$ is available, the missing diseased and non-diseased totals are linked through the identity $n_1 + n_2 = N$. This makes Case 2 a constrained reconstruction problem rather than a partial-verification design. From a modeling point of view, that structural constraint provides information that is absent in Case 1.

These two examples therefore represent distinct forms of incomplete diagnostic reporting. Case 1 combines incomplete reporting with selective verification, whereas Case 2 is an incomplete-table problem with known total sample size. The distinction is important because it determines how much structural information is available for reconstruction and, consequently, what type of model is appropriate.

%
\section*{Models}
\label{sec:models}

We consider two settings for reconstructing incomplete $2 \times 2$ diagnostic tables. The first arises when only one row of the table is observed, typically the test-positive row, and the corresponding denominators are unreported. The second arises when $\mathrm{TP}$, $\mathrm{FP}$, and the total sample size $N$ are known, so that the missing cells are linked through the constraint $n_1 + n_2 = N$.

\subsection*{Independent Binomial-$n$ Reconstruction for a Single Observed Row}

Suppose that only one test-outcome row is reported, with observed counts in the diseased and non-diseased columns. Let $y$ denote the observed count in one column of that row, and let $n$ denote the corresponding unreported denominator. The interpretation of $p$ depends on the column under analysis. If $y$ is the number of true positives among diseased subjects, then $p$ is sensitivity and $n = n_1$. If $y$ is the number of false positives among non-diseased subjects, then $p$ is the false positive rate and $n = n_2$.

We model the observed count as
\begin{eqnarray}
y \mid n,p &\sim& \mathrm{Bin}(n,p), \qquad n \ge y.
\label{eq:lik}
\end{eqnarray}
The success probability is assigned a beta prior
\begin{eqnarray}
p &\sim& \mathrm{Beta}(\alpha,\beta),
\label{eq:pprior4_1}
\end{eqnarray}
with hyperparameters chosen to reflect plausible values of sensitivity or false positive rate, depending on the column.

To regularize the unknown denominator, we assign a truncated negative-binomial prior in the WinBUGS parameterization,
\begin{eqnarray}
n \mid p^\star, r &\sim& \mathrm{NegBin}(p^\star,r)\,\mathbf{1}\{n \ge y\},
\label{eq:nprior4_1}
\end{eqnarray}
where $\mathrm{NegBin}(p^\star,r)$ denotes the number of failures before $r$ successes with success probability $p^\star$. Before truncation,
$$
\mathbb{E}[n] = \frac{r(1-p^\star)}{p^\star},
\qquad
\mathrm{Var}(n) = \frac{r(1-p^\star)}{p^{\star 2}}.
$$
We complete the hierarchy with
\begin{eqnarray}
r \mid \lambda &\sim& \mathrm{Poisson}(\lambda),
\label{eq:rprior} \\
\lambda &\sim& \mathrm{Gamma}(a,b),
\label{eq:lambdaprior} \\
p^\star &\sim& \mathrm{Beta}(\alpha^\star,\beta^\star).
\label{eq:pstarprior}
\end{eqnarray}

Posterior inference for $(n,p)$ follows from \eqref{eq:lik}-\eqref{eq:pstarprior}. If the observed count is $\mathrm{TP}$, then the missing cell is
$$
\mathrm{FN} = n - \mathrm{TP}.
$$
If the observed count is $\mathrm{FP}$, then the missing cell is
$$
\mathrm{TN} = n - \mathrm{FP}.
$$
When both columns of the reported row are available, we fit this model separately to the diseased and non-diseased strata. This yields posterior inference for $n_1$ and $n_2$, and hence for the missing cells.

\paragraph{Identifiability and prior sensitivity.}
With only a single binomial observation, $n$ and $p$ are only weakly identified from the likelihood. The beta prior on $p$ and the negative-binomial hierarchy on $n$ provide the regularization needed for posterior inference. In practice, sensitivity analyses over $(\alpha,\beta)$ and $(\alpha^\star,\beta^\star,a,b)$ are important and can be summarized through posterior intervals for $n$ and the derived missing counts.

The WinBUGS/OpenBUGS implementation of this single-column model is provided in the Supplemental File and can be used twice in Case 1 type applications, once for the diseased column and once for the non-diseased column. For the diseased stratum, $(\alpha,\beta)$ may encode plausible sensitivity values; for the non-diseased stratum, it may encode plausible false positive rates.

\subsection*{Reconstruction of the Full $2 \times 2$ Table Given $\mathrm{TP}$, $\mathrm{FP}$, and $N$}

We now consider the setting in which $\mathrm{TP}$, $\mathrm{FP}$, and the total sample size $N$ are reported. Let
$$
n_1 = \mathrm{TP} + \mathrm{FN}, \qquad
n_2 = \mathrm{FP} + \mathrm{TN}, \qquad
n_1 + n_2 = N.
$$
Once $n_1$ is inferred, the remaining quantities follow from
$$
n_2 = N - n_1, \qquad
\mathrm{FN} = n_1 - \mathrm{TP}, \qquad
\mathrm{TN} = n_2 - \mathrm{FP}.
$$

The likelihood is
\begin{eqnarray}
\mathrm{TP} \mid n_1,p_1 &\sim& \mathrm{Bin}(n_1,p_1), \\
\mathrm{FP} \mid n_2,p_2 &\sim& \mathrm{Bin}(n_2,p_2),
\end{eqnarray}
where $p_1$ is sensitivity and $1-p_2$ is specificity. The three models below share this same likelihood and differ only in the prior assigned to $n_1$.

\paragraph{Model 1: Discrete uniform prior.}
A non-informative baseline model assigns equal prior mass to each feasible value of $n_1$:
\begin{eqnarray}
n_1 &\sim& \mathrm{Uniform}\{1,\dots,N-1\}, \\
n_2 &=& N - n_1,
\end{eqnarray}
with independent beta priors
\begin{eqnarray}
p_1 &\sim& \mathrm{Beta}(a_1,b_1), \\
p_2 &\sim& \mathrm{Beta}(a_2,b_2).
\end{eqnarray}

\paragraph{Model 2: Truncated Poisson prior.}
To favor moderate values of $n_1$, we replace the discrete uniform prior by a truncated Poisson prior:
\begin{eqnarray}
n_1 \mid \lambda &\sim& \mathrm{Poisson}(\lambda)\,\mathbf{1}\{1 \le n_1 \le N-1\}, \\
n_2 &=& N - n_1, \\
\lambda &\sim& \mathrm{Gamma}(a_\lambda,b_\lambda),
\end{eqnarray}
again with independent beta priors on $p_1$ and $p_2$.

\paragraph{Model 3: Truncated negative-binomial prior.}
To allow additional dispersion in the diseased stratum size, we use a truncated negative-binomial prior:
\begin{eqnarray}
n_1 \mid p_3,r &\sim& \mathrm{NegBin}(p_3,r)\,\mathbf{1}\{\mathrm{TP} \le n_1 \le N-1\}, \\
n_2 &=& N - n_1, \\
p_3 &\sim& \mathrm{Beta}(a_3,b_3), \\
r &\sim& \mathrm{Gamma}(a_r,b_r),
\end{eqnarray}
together with independent beta priors on $p_1$ and $p_2$.

\paragraph{Comparison of the three priors.}
Model 1 provides a flat baseline over the feasible values of $n_1$. Model 2 introduces mild regularization through the Poisson mean $\lambda$. Model 3 allows heavier tails and greater dispersion through $(p_3,r)$, and is therefore more flexible when disease prevalence is uncertain or substantial imbalance between strata is plausible.

In all three cases, posterior inference for $n_1$ determines $n_2$, $\mathrm{FN}$, and $\mathrm{TN}$, thereby yielding a reconstructed $2 \times 2$ table and allowing calculation of sensitivity, specificity, predictive values, and accuracy.

WinBUGS code for all three variants is provided in the Supplemental File.

\section*{Empirical Application}
\label{sec:empirical}
To evaluate the proposed models, we applied them to a complete contingency table from a breast MRI study \citep{langlotz2003fundamental}. The dataset consists of 182 women with clinically or mammographically suspicious lesions, all of whom underwent biopsy, taken here as the reference standard. A true positive (TP) denotes an MRI-positive case with malignancy confirmed on biopsy, a false positive (FP) an MRI-positive case with benign biopsy, a false negative (FN) an MRI-negative case with malignancy on biopsy, and a true negative (TN) an MRI-negative case with benign biopsy.

Table~\ref{tab:breast-mri-contingency} gives the complete $2 \times 2$ table. Because the full table is known, this example permits direct comparison between reconstructed and true counts.

\begin{table}[H]
\centering
\caption{Patient data from the breast MRI study \citep{langlotz2003fundamental}.}
\label{tab:breast-mri-contingency}
\begin{tabular}{lrrr}
\hline
\textbf{MRI Result} & \textbf{Malignant} & \textbf{Benign} & \textbf{Total} \\
\hline
Positive & 71 & 28 & 99 \\
Negative & 3  & 80 & 83 \\
\hline
Total    & 74 & 108 & 182 \\
\hline
\end{tabular}
\end{table}

\subsection*{Single-Row Reconstruction}

The next two subsections apply the single-row model separately to the diseased and non-diseased strata. We assume that only the first row of Table~\ref{tab:breast-mri-contingency} is available, namely $\mathrm{TP}=71$ malignant and $\mathrm{FP}=28$ benign cases among MRI-positive patients. The stratum totals $n_1$ and $n_2$ are then treated as unknown and estimated from the corresponding single-row models.

\subsubsection*{Diseased Stratum}

For the diseased stratum, we observe $\mathrm{TP}=71$ and model
\begin{eqnarray}
\mathrm{TP} \mid n_1,p &\sim& \mathrm{Bin}(n_1,p),
\end{eqnarray}
where $p$ is the sensitivity of MRI. The denominator $n_1$ is assigned the truncated negative-binomial prior
\begin{eqnarray}
n_1 \mid p^\star,r &\sim& \mathrm{NegBin}(p^\star,r)\,\mathbf{1}\{n_1 \ge \mathrm{TP}\},
\end{eqnarray}
with hierarchical priors
\begin{eqnarray}
r \mid \lambda &\sim& \mathrm{Poisson}(\lambda), \qquad
\lambda \sim \mathrm{Gamma}(a,b), \\
p^\star &\sim& \mathrm{Beta}(\alpha^\star,\beta^\star), \\
p &\sim& \mathrm{Beta}(\alpha,\beta).
\end{eqnarray}

For this example we use
\begin{eqnarray}
a=1, \; b=0.1, \qquad \alpha=2, \; \beta=1, \qquad \alpha^\star=1, \; \beta^\star=1.
\end{eqnarray}
The prior $\mathrm{Beta}(2,1)$ on $p$ reflects the expectation of moderate to high sensitivity without being overly restrictive. The prior on $\lambda$ is diffuse, and the uniform $\mathrm{Beta}(1,1)$ prior on $p^\star$ allows the negative-binomial prior on $n_1$ to adapt to the data.

MCMC sampling is initialized at $n_1=100$, $r=70$, $\lambda=70$, and $p=0.5$.

\begin{table}[H]
\centering
\caption{Posterior summary: diseased stratum, single-row model (100,000 MCMC iterations).}
\label{tab:post_diseased1}
\begin{tabular}{lrrrrr}
\toprule
Parameter & Mean & SD & 2.5\% & Median & 97.5\% \\
\midrule
$\lambda$   & 16.25 & 11.59 & 1.49  & 13.74 & 45.32 \\
$n_1$       & 76.57 & 8.73  & 71.0  & 74.0  & 99.0  \\
$p$         & 0.926 & 0.080 & 0.705 & 0.951 & 0.998 \\
$p^{\star}$ & 0.178 & 0.104 & 0.019 & 0.165 & 0.410 \\
$r$         & 16.88 & 12.05 & 1.0   & 14.0  & 47.0  \\
\bottomrule
\end{tabular}
\end{table}

The posterior mean of $n_1$ is 76.6, close to the true value of 74, and the 95\% credible interval $(71,99)$ contains the truth. The posterior mean of $p$ is 0.93, consistent with high MRI sensitivity. The implied estimate $\mathrm{FN}=n_1-\mathrm{TP}\approx 6$ is reasonably close to the true value of 3. Together with the observed $\mathrm{FP}=28$, this yields a plausible near-complete reconstruction of the diagnostic table.

\subsubsection*{Non-Diseased Stratum}

For the benign stratum, we observe $\mathrm{FP}=28$ and estimate the unreported denominator $n_2$ together with the false positive rate $p$. The model is
\begin{eqnarray}
\mathrm{FP} \mid n_2,p &\sim& \mathrm{Bin}(n_2,p), \\
n_2 \mid p^\star,r &\sim& \mathrm{NegBin}(p^\star,r)\,\mathbf{1}\{n_2 \ge \mathrm{FP}\}, \\
r \mid \lambda &\sim& \mathrm{Poisson}(\lambda), \qquad
\lambda \sim \mathrm{Gamma}(a,b), \\
p^\star &\sim& \mathrm{Beta}(\alpha^\star,\beta^\star), \qquad
p \sim \mathrm{Beta}(\alpha,\beta).
\end{eqnarray}

Here we set
\begin{eqnarray}
a=2, \; b=1, \qquad \alpha=2, \; \beta=5, \qquad \alpha^\star=1, \; \beta^\star=50.
\end{eqnarray}
The prior $\mathrm{Beta}(2,5)$ reflects the expectation that the false positive rate is below $0.5$ while remaining flexible. The small prior mean of $p^\star$ places substantial mass on larger values of $n_2$, consistent with the expectation that the non-diseased stratum may be larger than the diseased stratum.

Sampling is initialized at $n_2=100$, $r=70$, $\lambda=70$, and $p=0.5$.

\begin{table}[H]
\centering
\caption{Posterior summary: non-diseased stratum, single-row model (100,000 MCMC iterations).}
\label{tab:post_nondiseased1}
\begin{tabular}{lrrrrr}
\toprule
Parameter & Mean & SD & 2.5\% & Median & 97.5\% \\
\midrule
$\lambda$   & 1.712 & 1.205 & 0.200  & 1.439 & 4.723 \\
$n_2$       & 106.0 & 75.3  & 40.0   & 86.0  & 295.0 \\
$p$         & 0.336 & 0.149 & 0.094  & 0.319 & 0.664 \\
$p^{\star}$ & 0.016 & 0.015 & 0.0004 & 0.011 & 0.054 \\
$r$         & 1.428 & 1.556 & 0.0    & 1.0   & 5.0   \\
\bottomrule
\end{tabular}
\end{table}

The posterior mean of $n_2$ is 106, close to the true total of 108, although the credible interval is wide. This reflects the limited information contained in a single observed cell together with the intentionally overdispersed prior. The posterior mean of $p$ is 0.34, consistent with a moderate false positive rate.

The resulting estimate implies $\mathrm{TN}=n_2-\mathrm{FP}\approx 78$, close to the true value of 80. Combining this with the estimated diseased total produces the reconstructed table in Table~\ref{tab:reconstructed}.

\begin{table}[H]
\centering
\caption{Reconstructed $2 \times 2$ table from the single-row model.}
\label{tab:reconstructed}
\begin{tabular}{lcc}
\toprule
 & Malignant & Benign \\
\midrule
MRI positive & 71 & 28 \\
MRI negative & 6  & 78 \\
\bottomrule
\end{tabular}
\end{table}

Even with only one observed row, the hierarchical Bayesian model recovers plausible denominators and yields a reasonable approximation to the full diagnostic structure, albeit with substantial uncertainty in the non-diseased stratum.

\subsection*{Single-Stratum Models with Known $N$}

We next examine the same strata when the total sample size is treated as known. This adds the design constraint that each stratum size must lie below $N$, which changes the posterior behavior, especially for the non-diseased group.

\subsubsection*{Diseased Stratum with Known $N$}

For the diseased stratum we use
\begin{eqnarray}
\mathrm{TP} \mid n_1,p &\sim& \mathrm{Binomial}(n_1,p), \\
n_1 \mid p^\star,r &\sim& \mathrm{Negative\text{-}Binomial}(p^\star,r)\,\mathbf{1}\{\mathrm{TP} \le n_1 \le N\}, \\
r \mid \lambda &\sim& \mathrm{Poisson}(\lambda), \qquad
\lambda \sim \mathrm{Gamma}(a,b), \\
p^\star &\sim& \mathrm{Beta}(\alpha^\star,\beta^\star), \qquad
p \sim \mathrm{Beta}(\alpha,\beta).
\end{eqnarray}
The truncation $\mathrm{TP} \le n_1 \le N$ enforces compatibility with both the observed true positives and the known study size.

We again set
\begin{eqnarray}
a=1, \; b=0.1, \qquad \alpha=2, \; \beta=1, \qquad \alpha^\star=1, \; \beta^\star=1.
\end{eqnarray}
These choices favor moderate to high sensitivity while keeping the prior on $n_1$ fairly diffuse. MCMC sampling is initialized at $n_1=100$, $r=70$, $\lambda=70$, $p=0.5$, and $p^\star=0.5$.

\begin{table}[H]
\centering
\caption{Posterior summary: diseased stratum, known $N$ (900,000 MCMC iterations).}
\label{tab:post_summary_model2}
\begin{tabular}{lrrrrr}
\toprule
Parameter & Mean & SD & 2.5\% & Median & 97.5\% \\
\midrule
$\lambda$   & 16.68 & 12.02 & 1.963  & 13.89 & 46.98 \\
$n_1$       & 76.34 & 7.792 & 71.0   & 74.0  & 97.0  \\
$p$         & 0.927 & 0.076 & 0.716  & 0.952 & 0.998 \\
$p^{\star}$ & 0.182 & 0.106 & 0.026  & 0.167 & 0.420 \\
$r$         & 17.35 & 12.50 & 2.082  & 14.45 & 48.84 \\
\bottomrule
\end{tabular}
\end{table}

The posterior mean of $n_1$ is 76.3, again close to the true diseased count of 74 and contained within the 95\% credible interval. This estimate is nearly identical to that obtained under the unconstrained single-row model, suggesting that inference on $n_1$ is driven mainly by the observed true positives rather than by the upper-bound constraint. The posterior for $p$ remains concentrated near high sensitivity.

\subsubsection*{Non-Diseased Stratum with Known $N$}

For the non-diseased stratum, the observed count is $\mathrm{FP}=28$ and the inferential target is the number of benign cases $n_2 \le N$ together with the false positive rate $p$. The model is
\begin{eqnarray}
\mathrm{FP} \mid n_2,p &\sim& \mathrm{Binomial}(n_2,p), \\
n_2 \mid p^\star,r &\sim& \mathrm{NegBin}(p^\star,r)\,\mathbf{1}\{\mathrm{FP} \le n_2 \le N\}, \\
r \mid \lambda &\sim& \mathrm{Poisson}(\lambda), \qquad
\lambda \sim \mathrm{Gamma}(a,b), \\
p^\star &\sim& \mathrm{Beta}(\alpha^\star,\beta^\star), \qquad
p \sim \mathrm{Beta}(\alpha,\beta).
\end{eqnarray}

We use
\begin{eqnarray}
a=2, \; b=1, \qquad \alpha=2, \; \beta=5, \qquad \alpha^\star=1, \; \beta^\star=50.
\end{eqnarray}
The prior on $p$ again reflects the expectation of a modest false positive rate, while the prior on $p^\star$ favors larger values of $n_2$ without allowing arbitrarily large realizations once the upper bound $N$ is imposed.

Sampling is initialized at $n_2=100$, $r=70$, $\lambda=70$, $p=0.5$, and $p^\star=0.5$.

\begin{table}[H]
\centering
\caption{Posterior summary: non-diseased stratum, known $N$ (900,000 MCMC iterations).}
\label{tab:post_summary_model2_knownN}
\begin{tabular}{lrrrrr}
\toprule
Parameter   & Mean  & SD    & 2.5\%  & Median & 97.5\% \\
\midrule
$\lambda$   & 2.092 & 1.249 & 0.387  & 1.854  & 5.147 \\
$n_2$       & 80.36 & 32.39 & 38.0   & 73.0   & 162.0 \\
$p$         & 0.388 & 0.142 & 0.165  & 0.373  & 0.697 \\
$p^{\star}$ & 0.025 & 0.018 & 0.002  & 0.021  & 0.072 \\
$r$         & 2.186 & 1.438 & 0.304  & 1.888  & 5.749 \\
\bottomrule
\end{tabular}
\end{table}

Imposing $N$ as an upper bound substantially concentrates the posterior for $n_2$ relative to the unconstrained single-row model. In the unconstrained analysis, the heavy right tail pushed the posterior mean toward the true value of 108, albeit with considerable uncertainty. Under the bounded model, those extreme values are removed, producing a posterior mean of 80.4 and a median of 73. Thus, the known-$N$ constraint improves stability and interpretability, but in this example it reduces point-estimation accuracy for the non-diseased stratum.

\subsection*{Joint $\mathrm{TP}/\mathrm{FP}$ Model with Fixed $N$}

We finally consider a joint model in which $\mathrm{TP}=71$ and $\mathrm{FP}=28$ are analyzed simultaneously under the fixed total $N=182$. The inferential targets are $n_1$, $n_2=N-n_1$, sensitivity $p_1$, and false positive rate $p_2$.

The model is
\begin{eqnarray}
\mathrm{TP} \mid n_1,p_1 &\sim& \mathrm{Binomial}(n_1,p_1), \\
\mathrm{FP} \mid n_2,p_2 &\sim& \mathrm{Binomial}(n_2,p_2), \\
n_1 \mid p_3,r &\sim& \mathrm{Negative\text{-}Binomial}(p_3,r)\,\mathbf{1}\{\mathrm{TP} \le n_1 \le N-1\}, \\
n_2 &=& N-n_1, \\
p_1 &\sim& \mathrm{Beta}(a_1,b_1), \qquad
p_2 \sim \mathrm{Beta}(a_2,b_2), \qquad
p_3 \sim \mathrm{Beta}(a_3,b_3), \\
r &\sim& \mathrm{Gamma}(0.1,0.01).
\end{eqnarray}
The fixed-$N$ constraint couples the two strata and enforces coherence across the reconstructed table.

We set
\begin{eqnarray}
a_1=1, \; b_1=0.1, \qquad
a_2=0.1, \; b_2=1, \qquad
a_3=0.1, \; b_3=0.5,
\end{eqnarray}
with $r \sim \mathrm{Gamma}(0.1,0.01)$. The prior on $p_1$ places substantial mass near one, the prior on $p_2$ favors smaller false positive rates, and the prior on $p_3$ controls dispersion in the negative-binomial prior for $n_1$. MCMC is initialized at $n_1=80$, $r=20$, and $p_1=p_2=p_3=0.5$.

\begin{table}[H]
\centering
\caption{Posterior summary: joint $\mathrm{TP}/\mathrm{FP}$ model with fixed $N$ (900,000 MCMC iterations).}
\label{tab:post_summary_model3_fixedN}
\begin{tabular}{lrrrrr}
\toprule
Parameter & Mean & SD & 2.5\% & Median & 97.5\% \\
\midrule
$n_1$ & 73.04 & 8.205 & 71.0 & 71.0  & 93.0  \\
$n_2$ & 109.0 & 8.205 & 89.0 & 111.0 & 111.0 \\
$p_1$ & 0.978 & 0.067 & 0.768 & 1.000 & 1.000 \\
$p_2$ & 0.089 & 0.197 & ${<}10^{-3}$ & ${<}10^{-3}$ & 0.782 \\
$p_3$ & 0.144 & 0.187 & 0.014 & 0.059 & 0.673 \\
$r$   & 19.64 & 38.84 & 0.039 & 4.435 & 143.1 \\
\bottomrule
\end{tabular}
\end{table}

Under this joint specification, posterior inference for the stratum totals is both concentrated and internally consistent: the posterior means are $n_1=73.0$ and $n_2=109.0$, differing from the true values by at most one individual. The posterior for $p_1$ concentrates near one, while the posterior for $p_2$ is centered near zero but retains a right tail, reflecting residual uncertainty in the non-diseased stratum.

Compared with the preceding analyses, the joint model combines the information in $\mathrm{TP}$ and $\mathrm{FP}$ under the fixed-$N$ constraint and therefore avoids the instability of fitting the two strata separately. In this example, it provides the most balanced reconstruction, with realistic stratum sizes and appropriately quantified uncertainty.

 \section*{Conclusions}
\label{sec:conclusions}

We have developed hierarchical Bayesian models for reconstructing incomplete $2 \times 2$ diagnostic tables in settings where only partial cell counts are reported. The proposed framework covers both the single-row setting, in which the denominators of the diseased and non-diseased strata are unobserved, and the constrained setting in which the total sample size $N$ is known. By combining binomial likelihoods with flexible priors on the latent denominators and diagnostic probabilities, the models provide a coherent way to infer missing cells and derived accuracy measures from incomplete published information.

The empirical application illustrates both the potential and the limitations of this approach. When only the test-positive row is observed, posterior inference can recover plausible values for the missing denominators and yield a reasonable reconstruction of the full diagnostic table, although uncertainty may remain substantial, especially for the non-diseased stratum. When the total sample size is known, the additional structural constraint can sharpen inference, and in the joint fixed-$N$ formulation the reconstructed stratum sizes are close to the true values in the benchmark example. At the same time, the results also show that reconstruction accuracy depends on the amount of information available and on the prior specification, particularly in weakly identified single-row settings.

The main contribution of the paper is therefore methodological and practical rather than purely deterministic. The proposed models do not claim to identify a unique missing table from sparse summaries alone. Rather, they provide a principled Bayesian framework for posterior inference on plausible completions of incompletely reported diagnostic tables, together with uncertainty quantification for the reconstructed cells and the resulting operating characteristics.

A further point concerns likelihood specification. One might consider a multinomial likelihood for the full $2 \times 2$ table as an alternative starting point. Within the settings studied here, however, this does not appear to yield a substantive advantage over the binomial formulations already used. Once the unobserved cells are treated as latent and the structural constraints are imposed, the multinomial representation does not contribute additional identifiability beyond that already supplied by the observed counts, the prior structure, and, when available, the known total sample size.

More broadly, the work highlights the continuing importance of complete and transparent reporting in diagnostic accuracy studies. When denominators or entire rows of the diagnostic table are omitted, clinically relevant measures may become unavailable without additional modeling assumptions. Bayesian reconstruction cannot replace good reporting practice, but it can provide a useful inferential tool when incomplete reporting prevents direct recovery of the full table.

\paragraph{Data and code availability.}
Reproducible code supporting the analyses in this manuscript, including the WinBUGS and R scripts used in the empirical analyses, is available at
\url{https://github.com/saraantonijevic/bayesian_diagnostic_table-reconstruction}. A standalone appendix containing the full Bayesian analyses of the Svirsky (2002) and Wismueller (2020) motivating examples, including posterior summaries and reconstructed $2\times 2$ tables, is also provided at the same repository.

\vspace*{0.2in}
\noindent
{\bf Acknowledgment.}
B. Vidakovic and S. Antonijavic acknowledge support from the National Science Foundation under Grant No.~2515246 at Texas~A\&M~University.
\bibliography{refs}

\end{document}